\documentclass[prb,aps,groupedaddress,amsmath,twocolumn,showpacs]{revtex4}

\usepackage[dvips]{graphicx}
\usepackage{dcolumn}
\usepackage{bm}
\usepackage[version=3]{mhchem} % Formula subscripts using \ce{}

\begin{document}

%%%%%%%%%%%%%%%%%%

%\widetext

%\twocolumn
%[\hsize\textwidth\columnwidth\hsize\csname
%@twocolumnfalse\endcsname
%%%%%%%%%%%%%%%%%%

\title{ Linearized Auxiliary fields Monte Carlo: efficient sampling of the fermion sign} 
\author{Sandro Sorella}
\email[]{sorella@sissa.it}
\affiliation{SISSA -- International School for Advanced Studies, Via Bonomea 265
  34136 Trieste, Italy} \affiliation{Democritos Simulation Center
  CNR--IOM Istituto Officina dei Materiali, 34151 Trieste, Italy} 
\date{\today}

\begin{abstract}
We introduce a method  that combines the power of both the lattice Green function Monte Carlo (LGFMC)  with the auxiliary field techniques (AFQMC), 
and allows us to compute 
{\em exact} ground state properties of the Hubbard model for $U\lesssim 4t$ 
on finite clusters.
 Thanks to LGFMC one obtains unbiased zero temperature results,  not 
affected by the so called Trotter approximation of the imaginary time propagator $e^{ - H \tau}$. On the other hand the AFQMC formalism yields 
a remarkably fast convergence in $\tau$ before the fermion sign 
problem becomes prohibitive.  
As a first application we report ground state energies  
 in the Hubbard model at $U/t=4$ with up to one hundred sites. 
   
\end{abstract}
\maketitle
After several years of scientific 
effort, based on advanced analitycal and numerical methods, only 
very few properties of the 2D Hubbard model have been settled.
The 2D Hubbard model is defined in a 
square lattice  containing a finite number $L$ ($N$) of sites (electrons):
\begin{equation}
H = -t \sum_{<i,j>,\sigma} c^{\dag}_{i,\sigma} c_{j,\sigma} + U \sum_i n_{i}^\uparrow n_i^\downarrow
\end{equation}
with standard notations. 
In the thermodinamic limit, namely for   $L\to \infty$ 
at given density $\rho=N/L$ fundamental issues  
such as the existence of a ferromagnetic phase at large 
$U/t$ ratio and/or the stability of an  homogeneous ground state
with possible d-wave supoerconducting  properties  are 
still  highly debated,  as several approximate numerical techniques  lead 
to controversial and often conflicting  results. 
This situation is particularly embarazzing, since  
recent progress in the realization of fermionic optical lattices could lead 
soon to the experimental realization of the fermionic Hubbard model,
apparently much before we will reach a consensus among the different 
theoretical and numerical techniques.

{\rm Method:} In LGFMC\cite{ldmc}, 
the main property used  
to compute the ground state $\psi_0$
 of a many body Hamiltonian $H$, is 
the iterative  application of a {\em linearized} Green's function:
\begin{equation} \label{lgreen}
G= \Lambda I - { H } 
\end{equation}
to an initial wave function $\psi$ by a stochastic method, 
namely $\psi_0 = \lim\limits_{n\to \infty} G^n | \psi \rangle $.  
Here $I$ is the identity matrix and  $\Lambda$ is a suitably large constant.
This is possible because 
the application of $G$ to a given configuration
$|x\rangle $, where electrons have definite positions and spins, 
can be expressed as a sum of a finite number of 
independent configurations. Namely the number of non zero matrix elements 
$ G_{x^\prime,x} = \langle x^\prime | G | x \rangle$ for given $x$ is 
affordable ($\propto L$), though the Hilbert space is exponentially 
large with $L$.
For large $n$  it is possible to sample the ground state wave function
$\psi_0(x) = \langle x | \psi_0 \rangle $ and its correlation functions.
This method can be improved by the so called importance sampling, yielding  
a much more efficient algorithm: 
the matrix elements of $G$  are scaled by means 
of the so called guiding function $\psi_g(x)$:
\begin{equation}
 \bar G_{x^\prime,x}  = \psi_g(x^\prime) G_{x^\prime,x} /\psi_g(x)
\end{equation}
This method allows the calculation of exact ground state properties without 
any approximation other than {\em (i)}  
 {\em  finite} $L$ and 
{(ii)} statistical errors. The latter  
may be particularly large with fermionic systems, 
due to the unfamous ''sign problem'', a limitation 
that is usually much severe, if not prohibitive,  for this type of approach.
We remark also that, in LGFMC,
one can work with infinite $\Lambda$\cite{caprio}
and sample  the many body propagator $e^{-H \tau}= \lim \limits_{\Lambda  \to \infty} ({G\over \Lambda})^{n} $ with a similar computational effort, despite $n= \tau \Lambda \to \infty$ in this limit.
Moreover, since the projection time $\tau$ is proportional to the length of the 
 Monte Carlo simulation, converged ground state properties are easily  
obtained after the equilibration time, and  
the limit $\tau \to \infty$ is basically achieved 
without any particular effort when there is no sign problem.     
% as the 
%statistical errors grow exponentially with $n$.

Another important stochastic method that is quite popular for the 
Hubbard model is based on 
  AFQMC\cite{hirsch,canieporci} 
and the algebra of one 
body propagators. 
The basic property used in this approach is that 
a one body operator $U$, when  applied 
to a Slater determinant $|D\rangle$,
generates again a Slater determinant $| D^\prime \rangle = U|  D \rangle $, that is easy to evaluate. 
Obviously, by the above definition, the product of several  one-body 
propagators remains a one-body propagator, and the computation is 
always feasible for a finite number of them. 
Indeed, within AFQMC, the many body propagator $e^{- H \tau}$ 
 is conveniently  written as a superposition of time dependent 
one body propagators 
$U_{[\sigma]}( \tau, 0) = U_\sigma(t_{L_\tau})  
\cdots  U_\sigma(t_1 )  $
after the  introduction of discrete Hubbard-Stratonovich 
time-dependent auxiliary fields $\sigma_i (t_j) = \pm 1 $ defined in all the $L$ lattice sites 
and a finite number of  discrete imaginary times $t_j= j \tau/L_\tau$ 
$j=1,\cdots, L_\tau$.
Thus $ e^{-H \tau } | \psi \rangle = \sum\limits_{\sigma}  
 U_\sigma (\tau,0) |\psi \rangle $ 
for large imaginary time gives formally the exact ground state wave function 
and can be computed by Monte Carlo sampling 
of the auxiliary fields $\sigma_i (t_j) =\pm 1$.\cite{hirsch}
At variance of  the LGFMC technique, 
it is not simple to avoid the error due to the 
discretization in time of the propagator-usually called Trotter error-
and all the results require a careful and often boring extrapolation both in 
$\tau \to \infty$ and $L_\tau \to \infty$.
By contrast, within AFQMC, the sign problem is much less severe 
for two main reasons, (i)  for $U=0$ the method is exact and one does not need any random sampling, (ii) at half-filling or with negative $U$ 
 the method is not affected by the ''sign problem''.  

In the following we propose to combine the power of the two methods, 
by taking the best of the two approaches in what we name  
Linearized   Auxiliary fields Monte Carlo (LAQMC). 
From AFQMC we take the advantage 
of a much less severe sign problem and from LGFMC  
the exact imaginary time projection 
will be available  in a rigorous and simple way.
The latter  achievement has been made recently 
 possible also  within the so called 
''diagrammatic'' Monte Carlo\cite{prokovev,lichtnestein}, but only within the 
reasonable assumption that the diagrammatic perturbation serie converges.   
%in the   method we propose we  remove also the limitation of working with 
%a finite projection time $\tau$, taking the 
%advantage of the GFMC scheme, as discussed before. 

In order to define this new approach we use that the lattice 
Green function (\ref{lgreen}) of the Hubbard model 
can be  written {\em exactly}  as a {\em finite} sum of 
one body propagators $U_i$, with an approach that is similar 
to the conventional auxiliary field technique, where instead of splitting up
the many propagator $\exp(-\tau H)$ we focus on its linearized 
expression given by $G$:
\begin{equation} \label{greenhst}
G = \Lambda I - H =\sum_{i=1}^{p}  a_i   U_i
\end{equation}
where 
the coefficients $a_i \ge 0 $ and the $U_i$ will be defined later on and $p=4L$ 
for the Hubbard model.
The above identity can be  generally fullfilled  
for {\em any} reasonable many-body Hamiltonian by using a number $p$ of 
one-body propagators that scales at most as $L^2$. 
For the Hubbard model we obtain after simple algebra: 
\begin{eqnarray}
\Lambda &=&  \Gamma_U ( e^\lambda+ e^{-\lambda} -2) N+\sum\limits_{i=1}^p a_i\nonumber \\
U_i &=&  e^{ - \gamma_i c^{\dag}_{k_i,\sigma_i} c_{k_i,\sigma_i} } ~~ i=1,\cdots 2 L \nonumber \\
U_{i+5/2 L \pm L/2 } & = &e^{ \pm  \lambda ( n_{i}^{\uparrow} - n_{i}^{\downarrow} ) } 
~~ i = 1,\cdots L \nonumber 
 \end{eqnarray}
where, in order to satisfy Eq.(\ref{greenhst}):
\begin{eqnarray}
a_i ( e^{-\gamma_i}- 1) &=& -\epsilon_{k_i}  \label{kinhst}  ~~~ i \le 2L  \\
2 a_i  (e^{\lambda} -1) ( e^{-\lambda} -1) &=& -U     ~~~~ i > 2L 
\end{eqnarray}
and  $\epsilon_{k^i}=-2t (\cos(k_x^i) + \cos(k_y^i))$ is the $U=0$ 
band and $k^i$ label all the $2L$ independent  $k$ vectors 
of the spin-up and spin-down electrons ($\sigma_i=\uparrow$ for 
$i=1,\cdots L$ and $\sigma_i=\downarrow$ for $i=L+1,\cdots 2 L$).
Here for simplicity we set $a_i=\Gamma_t$ ($a_i=\Gamma_U$), 
independent of $i$ for $i \le 2L$ ($i>2L$). 
Notice that the main difference compared with the discrete Hubbard-Stratonovich transformation\cite{hirsch} is that, in order to decompose the propagator $G$, 
we introduce not only one body propagators for the 
interaction term $U$, but we use also further  
one body operators $U_i$ for $i \le 2L$, to recast the kinetic 
term as a simple sum of one-body propagators. 
In some sense this is equivalent to double the dimension
of the auxiliary fields $[\sigma]$, extension that  does not lead to 
a particular loss of efficiency 
of the algorithm and, on the other hand, allows us to remove the bias due 
to the time discretization in a simple and rigorous way.
The choice of $\Gamma_t$ 
and $\Gamma_U$, and  in principle all the coefficients $a_i$, 
 are completely arbitrary in this 
approach and can be tuned for optimizing efficiency, by a substantial
alleviation of the sign problem.  
On the other hand, it is simple to realize that  
for $ \Gamma_t = 1/\Delta t$ and $\lambda = \sqrt{ U \Delta t} $, 
one recover the same Master equation\cite{hamman}
 of the standard AFQMC in the limit 
$\Delta t \to 0$, where $\Delta t $ is the time discretization adopted with 
the Trotter approximation.  
Thus at least in this limit  the proposed method has no sign problem in 
all the cases when the standard AFQMC has no sign problem.
We will refer as the ''time continuous limit'' (TCL)
for this particular choice of the coupling $\Gamma_t$ and $\lambda$.

Once this decomposition is implemented 
we immediately recover the main property of GFMC, provided the 
configuration $|x\rangle $ is replaced by a 
Slater determinant $| D \rangle$, defined by the  
orbital matrix $D_{ij}$, that in the particular case of the Hubbard model 
reads: 
\begin{equation} \label{formd}
| D \rangle = \prod\limits_{i=1}^{N^\uparrow} ( \sum_j D_{i,j} c^{\dag}_{j,\uparrow} ) \prod \limits_{i=N^\uparrow+1}^{N}  ( \sum_j D_{i,j} c^{\dag}_{j,\downarrow} ) | 0 \rangle 
\end{equation}
where $D$ is real and  $N^\uparrow$ is the number of spin-up particles.
In fact,  
the application of $G$, written in the form (\ref{greenhst}), to a single Slater determinant  generates a finite number of Slater determinants of the 
same form.
The importance sampling can be analogously defined by means of 
a guiding function $\psi_g : D \to \psi_g(D)$ such that 
$ \psi_g (D) $ can be easily computed. 
From this point of view the method is similar to Constrained Path AFQMC
\cite{cpqmc}, where the guiding function is  defined in terms 
of a simple mean field Slater determinant $|\psi_{MF}\rangle$, namely
$ \psi_g(D) = \langle \psi_{MF} | D \rangle $.
%Since in the proposed  method the one body-operators  can be taken generally 
%far from the identity, producing generally finite jumps from one determinant 
%$D$ to another one $U_i | D\rangle$, it is not necessary to introduce 
%smsp0201.819713.0earing of the guiding function to avoid lack of ergodicity.
%For this reason we have found that a very effective choice is to take 
%$\Gamma_t $ large enough, to have less sign problem,  by keeping fixed  
% $e^\lambda =20$. 

%However in order to sample correctly the exact propagation, here the mean-field
%state $ \psi_{MF} $ is not necessarily given by Eq.(\ref{formd}), 
%and, more importantly,  
% the resulting guiding 
%function is not allowed  to vanish exactly when the 
%determinant $D$ becomes 
%orthogonal to $| \psi_{MF}\rangle$. 

In order to work with a rigorous statistical method 
with {\rm finite variance}   
one can employ standard smearing procedures 
of the guiding function based on the reweighting method, that allow us to work 
with a non zero guiding function for all $D$ generated by the 
Markov process\cite{ceperleyrelease}. 
%that allow us to work with {\em finite variance} for all physical quantities, 
%even when the QMC sampling generate determinants $D$ almost orthogonal to 
%$\psi_{MF}$.
Following the argument discussed in Ref.\onlinecite{hydrogen}, a very 
efficient smearing procedure is defined here by means of the Green's function,
a $2L\times 2L$ matrix:
\begin{equation}
g_{i\sigma,j\sigma^\prime} = {\langle \psi_{MF}| c^{\dag}_{i,\sigma} c_{j,\sigma^\prime} | D \rangle \over \langle \psi_{MF} | D \rangle } 
\end{equation}
If the determinant $ |D | =\langle \psi_{MF} | D $ vanishes, some of the 
the Green's function elements should diverge in the same way, so that 
we define a  reweighting factor  $R \le 1 $, satisfying 
 $ R \simeq | D | $  for $ |D| \to 0$:
\begin{equation} \label{regularization}
R(D) = 1 / \sqrt{ 1 + \epsilon^2 Tr (g g^T)}
\end{equation}
where the trace $Tr (g g^T) = \sum_{i,j,\sigma,\sigma^\prime} 
|g_{i\sigma,j\sigma^\prime}|^2$, and $\epsilon$ is an appropriate small number, 
chosen in a way that, on   average,  $R$ is around $0.9 \div 0.95$.
Then it is easy to define a guiding function 
$\psi_g(D) ={ |D| \over R}$ that 
remains finite for $ |D| \to 0$. On the other 
hand  for $\epsilon$ 
small enough $\psi_g$  
remains sufficiently close to the mean field determinant $|D|$, by allowing  
a very good importance sampling.  

Let us consider the basic step of the stochastic implementation 
of the power method  $\Lambda -H$.
Once the normalization $b_D = \sum_{i=1}^p | \bar G_{U_i |D\rangle ,| D \rangle}| $
is given, the walker weight and the determinant $|D\rangle $ 
are updated by means of the following simple Markov chain:
\begin{eqnarray} \label{algorithm}
% e_n &\to&  e_n + e_L(D) R(D) \,w \, (C-1)/(b_D-E)  \nonumber \\ 
% e_d &\to&  e_d + R(D)\, w\, (C-1)/(b_D-E) \nonumber \\ 
 |D\rangle  & \to & U_i |D\rangle  {\rm ~~with ~ probability~~} | \bar G_{U_i | D\rangle ,| D\rangle }|/b_D \nonumber  \\
 w  & \to & w \, b_D \, {\rm Sgn} \bar G_{U_i |D\rangle ,| D\rangle } \label{chsign}
\end{eqnarray}
On the other hand the expectation value of the energy as any other 
''mixed average'' quantity, can be computed by means of the ratio of 
two random variable averages $ { < e_L(D)  R w > \over <  R w  > } $, where 
$e_L(D)$ is the local energy:
$e_L(D) = \Lambda -\sum_i a_i {  \langle \psi_{MF} | U_i | D \rangle \over \langle \psi_{MF} | D \rangle }$. 

Several walkers with weights $w_i, i=1,\cdots M$ 
evolve with the above Markov chain and undergo a 
''branching'' process\cite{calandra} to optimize efficiency in the 
sampling. 
%This reconfiguration process 
%is exact statistically, and allows us  
%to eliminate walkers with accumulated small weights, against the  
%survived ones  with larger weight, that in turn are appropriately replicated.
%After that all the walkers weights are set to one. 
In order to eliminate completely the finite population bias 
it is necessary  only to  bookkeep
  a ''correcting factor'' $\bar w ={ 1\over M}  \sum_i w_i$.\cite{calandra} 

{\bf Constrained path as a standard LGFMC:} 
In this formalism it is very simple
to employ the constrained path approximation\cite{cpqmc}. 
This is a very stable algorithm as 
the walker is constrained to avoid regions with extremely small 
determinant $|D|$. 
%We remark here that this approximation is exact when there is no sign problem 
%for appropriate choice of the guiding function, namely 
%when the mean-field determinant $|D|>0$ strictly. This is achieved by the 
%CPQMC in the 
%limit $\Gamma_t \to \infty$ (and in particular in the TCL), 
%where according to the mentioned master 
%equation there is no sign problem in all cases where the HST has no 
%sign problem. 
A simple  way to implement the CPQMC  approximation in the 
continuous limit is  obtained by the following standard recipe:  
%''killing'' the walkers that cross the nodal region:   
whenever a sign change occur  in Eq.(\ref{chsign}), 
the walker weight  $w$ is  simply annihilated within this approximation. 
 This implementation is particularly important for applying the 
standard release node\cite{ceperleyrelease}, explained below.
\begin{figure}%[h]
\centering
\includegraphics[width=0.80\columnwidth]{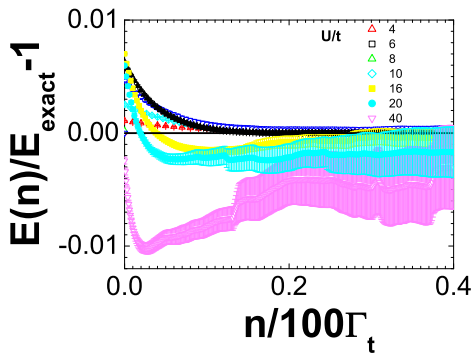}
\caption{
\label{qmcvsexact}
(Colors online) Comparison of exact ground state energy and 
the one obtained by LAQMC  projection for the Hubbard 
with  two holes ($N=16$) in the $18$ square lattice with $\Gamma_t=2.5 U$. 
Notice the rapid convergence with projection time even for large $U/t$.
} 
\end{figure}
 
{\bf Release node technique: }
When there is sign problem, the Markov chain (\ref{algorithm}) 
 is unstable because after 
a while half of the walkers will have negative sign and will cancel almost 
exactly the contribution of the ones with positive sign. 
In order to stabilize the process, we 
employ the standard release node technique introduced  long 
time ago\cite{ceperleyrelease}, and adapted to the present case, 
to take into account only a ''discrete time'' projection given by the 
power method.
%To each walker we  assign an integer $n_w$ that is initialized to zero.
%At each Markov step (\ref{algorithm}) this number $n_w$  is increased by one, but only after that the walker weight $w$ has changed its sign 
%for the first time.
%Then the weight $w$ is set to zero if $n_w> n_\tau$.
%For $n_w$ small enough, the branching process allows to remain with 
%a finite population of walkers that have never changed their signs, 
%namely with $n_w=0$. This population  can be considered an equilibrated 
%state representing the CPQMC approximate initial state. 
%The key idea of the release node algorithm is that 
%when the walker undergoes a Markov step (\ref{algorithm}) that thus not 
%change the sign of $w$ it  can be considered an evolution of a  
%stochastic state, propagated    
%either with the constrained path algorithm or with the 
%exact propagation.
Therefore after the Markov chain equilibrates, we can have access with a single 
run and a very simple postprocessing, to  all the history 
evolution of the energy as a function of the power 
method iterations 
starting from the very good estimate provided by the CPQMC state $\psi_{CPQMC}$, namely:
\begin{equation} \label{release}
 E_n = { \langle \psi_{MF} | H G^n | \psi_{CPQMC} \rangle \over 
         \langle \psi_{MF} |  G^n | \psi_{CPQMC} \rangle } 
\end{equation}
for all $n\le n_\tau$, where $n_\tau$ is the maximum release time allowed. 
%Indeed, after equilibration, 
%all the walkers with $n_w\le n$ contribute 
%to $E_n$ because
% if a walker has $n_w< n$ it can be considered to have 
%experienced further $n-n_w$ iterations with the exact Green function, 
%that, in absence of sign changes in Eq.(\ref{algorithm}), coincides with
%the one used to generate the equilibrated $CPQMC$ state. 

{\bf Guiding function:} In order to define the    
appropriate  guiding function $\psi_g$ 
at one electron per site filling $N=L$
we use an antiferromagnetic 
mean field Slater determinant 
with the order parameter along the x-spin axis\cite{becca,lugas}.
\begin{figure}%[h]
\centering
\includegraphics[width=1.0\columnwidth]{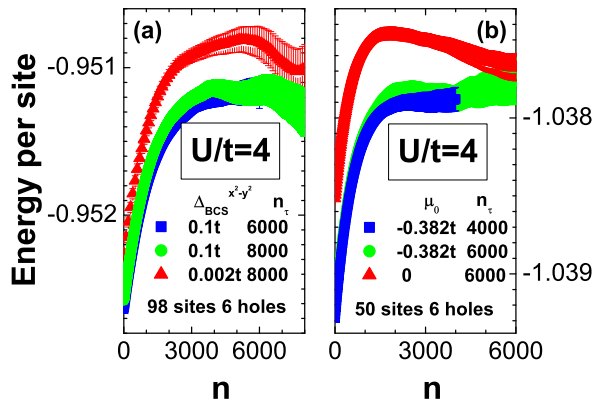}
\caption{
\label{test50}
(Colors online)   Energy as a function of  
the power method iterations using  different mean field wave functions
 $\psi_{MF}$  for defining the guiding function in LAQMC. 
They differ  for the value of the chemical potential $\mu_0$ 
and the BCS d-wave parameter $\Delta_{BCS}^{x^2-y^2}$ 
(see Ref.\onlinecite{epaps}). 
(a) $\mu_0=-0.198t$ (b) $\Delta_{BCS}^{x^2-y^2}=0.002t$}
%Each curve is obtained  
%by a single 
%simulation at given $n_{\tau}$ (see text)} 
%(b)  Plot of the average sign as a function of 
%the imaginary time projection, when  applied to the {\em same}   
% ''biased'' wave function obtained without sampling the sign.
%This projection is 
%obtained with the continuous time projection method\cite{caprio} for 
%the LAQMC  technique, as discussed in the text.
%} 
\end{figure}

Moreover away from half-filling we use also a BCS like wave function with 
a small d-wave order parameter $\Delta_{BCS}^{x^2-y^2}$, when the ground state of the $U=0$ model is degenerate (open shell case).  
In the half-filled case $N=L$ in any bipartite lattice 
and for  $U=0$, 
there is no sign problem 
(for $\Gamma_t$ large enough) and $\psi_{CPQMC} =\psi_0$. Therefore 
the method is exact already for $n=0$.
%Indeed  in our discretized implementation of the propagator,
%by keeping $\lambda$ fixed and employing the limit $\Gamma_t \to \infty$
%we obtain always the {\em exact} result within the constrained path 
%approximation, 
%because the probability to cross 
%from negative to positive sign region remains finite. 
%Moreover, in the latter 
%case,   a mean field state with a finite antiferromagnetic order parameter
%does never vanish for all determinant $D$ generated by the random walk and 
%therefore it is not necessary to use any regularization in the guiding 
%function to obtain exact results in bipartite lattice at half-filling 
%(similarly for negative $U$, it is enough to use a superconducting 
%mean field state with on site s-wave order). 
On the contrary the TCL will lead to biased result without smearing 
the guiding function ($\epsilon=0$), 
due to lack of 
ergodicity in the Markov process.
This  is  a well known problem  
in the standard CPQMC\cite{cpqmc},  that can be 
definitively solved  
by means of 
the careful regularization introduced for $\epsilon> 0$. 
As a particular example we show in Fig.(\ref{qmcvsexact}) 
the comparison of the exact results\cite{becca} for the two hole case 
in the 18 site cluster. 
%In this case we employed the continuous time 
%version of the algorithm in order to emphasize the rapid convergence of 
%the method, due to the high quality of the constrain path approximation. 
In this picture we reach convergence within statistical errors 
always with  a small number of power iterations  
with no particular difficulty to sample the sign even 
for large $U/t$ values.
From this picture we remark that 
only for very large  $U/t$ ratios we observe that  
the convergence to the exact result is non monotonic as a function of   
 $n$.
In Fig.~(\ref{test50})   the method is shown to work quite effectively as 
the CPQMC estimate remains very accurate also for a large 
cluster size, and convergence can be achieved much more quickly than the standard AFQMC\cite{hlubina}. 
In this figure we see that even when the initial wave function is not optimal
it is possible to reach convergence with a quite good error bar.
However it is also clear that a good initial guess allows 
 a much more accurate energy estimate by using a much smaller $n_{\tau}$.
 
%Moreover we see in Fig.2(a) how effective is to work with finite $\lambda$
%as we get a clear improvement of the average sign that is about 
%an order of magnitude larger at $\tau t=1$ compared to the standard AFQMC. 
%A finite large $\lambda$  (here we have chosen $e^\lambda=20$) 
%allows the walker 
% to escape more easily 
%from a region very close to the nodal surface $\psi_{g}(D)=0$, fact 
%that improves 
%the quality of the sampling, and therefore the average sign increases. 

\begin{figure}%[h]
\centering
\includegraphics[width=0.9\columnwidth]{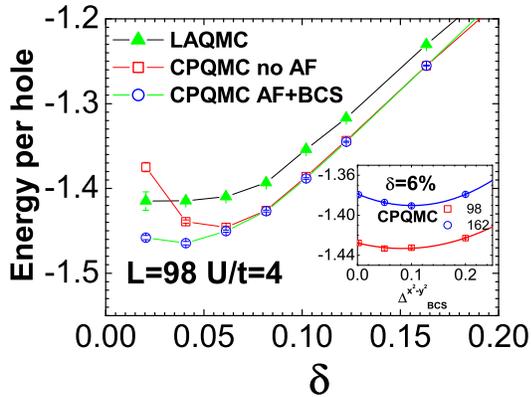}
\caption{
\label{phase}
(Colors online)  
Energy per hole for LAQMC release nodes (full symbols) 
and CPQMC (empty symbols). Lines are guides to the eye. 
} 
\end{figure}

Finally we show the results\cite{epaps} obtained for the energy per hole in Fig.(\ref{phase}), where $\delta=1-N/L$ is the hole concentration.
If there is phase separation between an hole rich phase and 
the undoped insulator at $\delta=0$
the energy per hole should have a minimum 
at a critical doping $\delta_c$\cite{emery}.
Our reference energy at $\delta=0$ is {\em exact}
 and given by $E/L= -0.85996(5)t$ 
for $L\to \infty$ at $U/t=4$.
For small   doping,  
%it is difficult to obtain converged  results for large size 
%starting from an homogeneous guiding function, and therefore 
though  we obtained a non monotonic 
behavior similar to the ones shown in  
Fig.~\ref{qmcvsexact} for $U/t>12t$, 
we were able to achieve convergence 
even in these particularly difficult cases using up to $n_{\tau} = 10000$
power method iterations.
%, where an horizontal energy per hole is expected 

%below $\delta_c$ as the Maxwell  construction provides a variational

%energy consistent with that in the thermodynamice limit. 
Considering therefore our energy results  
for the larger possible cluster size, we find  
a clear  flat region  in the energy per hole $e_h(\delta)$, as shown in 
Fig.(\ref{phase}), that may suggest an incipient phase separation\cite{jarrell}.
This behavior is in disagreement  with the 
 standard CPQMC, where  a clear minimum was found  
at doping $\delta_c \simeq 5\% $\cite{zhang} for $L\simeq 100$. 
In order to understand this result, we have performed CPQMC 
calculations with our technique, and found, as clearly displayed 
in the same picture,  that the 
CPQMC energy per hole is quite sensitive to the quality of the guiding 
function in the small doping region, and a magnetic guiding 
function containing both an antiferromagnetic and a d-wave order parameter 
greatly improves the CPQMC energy per hole, that becomes qualitatively 
similar to the LAQMC $e_h(\delta)$.
Summirizing  an  infinite compressibility appears plausible only 
when the doping approaches 
zero, in contrast with the clearly bounded one 
obtained in the $t-J$ model 
with a similar guiding function\cite{lugas}. 
We believe that this difference is due to 
the particle-hole symmetry, that is  not satisfied in 
the $t-J$ model, and instead could  
imply  $e_h(\delta) \simeq \delta^2$ as in 1D.

%At this stage we can safely  conclude that the charge compressibility 
%should be quite large at small doping, that could be in agreement with  
%the hyperscaling hupothesis\cite{imada}, and recent DMFT results\cite{jarrell}.

%Moreover, as shown in the inset,  we have found a clear energy gain  
%(within CPQMC) and a more stable projection (see Fig.\ref{test50}a) 
% with a sizabale d-wave BCS pairing  
%suggesting that the 
%$d-$wave superconding order can be stabilized at small doping, due to the 
%large charge compressibility.\cite{castellani}

In conclusion we have presented a new method for the simulation of  
the Hubbard model with about $100$ electrons, 
even when the sign problem prevents 
affordable calculations with standard techniques.
Preliminary results show a very large but finite compressibility
at small  doping in the square lattice Hubbard model at $U/t=4$.

This work was  supported by a  PRACE grant 2010PA0447.

%Unused bibitems

%Unused bibitems

%\bibitem{basis_benzene} S. Azadi, C. Cavazzoni, and S. Sorella,
%  Phys. Rev. B {\bf 82}, 125112 (2010). 


\begin{thebibliography}{99}
\bibitem{ldmc} 
 N. Trivedi and D. M. Ceperley  \prb {\bf 41}, 4552 (1990).

\bibitem{caprio} S. Sorella and L. Capriotti \prb {\bf 61}, 2599 (2000).
 

\bibitem{hirsch} J. E. Hirsch \prb {\bf 31}, 4403.

\bibitem{canieporci} S.R. White {\em et al} \prb {\bf 40}, 506, (1989); 
S. Sorella {\em et al.} Europhys. Lett. {\bf 8}, 663 (1989).  

\bibitem{prokovev} Philipp Werner {\it et al.} \prl {\bf  97}, 076405, (2006);
N. V. Prokof’ev and B. V. Svistunov \prb {\bf  77}, 125101 (2008).

\bibitem{lichtnestein} A. N. Rubtsov, V. V. Savkin, and A. I. Lichtenstein, 
\prb {\bf 72}, 035122 (2005). 

\bibitem{hamman} S. Fahy and D.R. Hamann \prb {\bf 43}, 765 (1991). 

\bibitem{cpqmc} S. Zhang, J. Carlson, and J.E. Gubernatis \prl {\bf 78}, 4486, 
(1997); J. Carlson {\it et al.}, \prb {\bf 59}, 12788 (1999). 

\bibitem{ceperleyrelease} D. M. Ceperley, and B. J. Alder, 
Journal of Chemical Physics, {\bf 81}, 5833 (1984).

\bibitem{hydrogen} C. Attaccalite and S. Sorella \prl {\bf 100}, 114501 (2008).


\bibitem{calandra} M. Calandra and S. Sorella, \prb   {\bf 57}, 11446 (1998).

\bibitem{becca} F. Becca, M. Capone  and S. Sorella \prb  {\bf 62}, 12700 (2000)

\bibitem{lugas} M. Lugas, L. Spanu, F. Becca, S. Sorella \prb {\bf 74}, 165122 
(2006).

\bibitem{hlubina} R. Hlubina , S. Sorella and F. Guinea,  \prl  {\bf 78}, 1343 (1997).

\bibitem{epaps} See Supplemental  Material at [URL] for energy values and finite size scaling at $U/t=4$.

\bibitem{emery} V. J. Emery, S. A. Kivelson, and H. Q. Lin, \prl  {\bf 64}, 475 (1990).

\bibitem{zhang} Chia-Chen Chang and Shiwei Zhang 
Phys. Rev. B 78, 165101 (2008).

\bibitem{jarrell} E. Khatami {\it et al.} \prb {\bf 81}, 201101 (2010).





%\bibitem{muramatsu} Z. Y. Meng, T. C. Lang, S. Wessel,

%  F. F. Assaad, and A. Muramatsu, Nature {\bf 464}, 847 (2010).

%D. F. B. ten Haaf, J. M.J. van Leeuwen, W. van Saarloos, and D. M. Ceperley,

%\prb {\bf 51} ,13039 {1995}.

%\bibitem{spanu_graphite} L. Spanu, S. Sorella, and G. Galli,

%  Phys. Rev. Lett. {\bf 103}, 196401 (2009). 

%\bibitem{lrdmc} M. Casula, C. Filippi, and S. Sorella, \prl {\bf 95}, 100201͑(2005); M. Casula, S. Moroni, S. Sorella and C. Filippi, J. Chem. Phys. 

%{\bf 132}, 154113 (2010). 

%\bibitem{sorella} S. Sorella \prb {\bf 64} 024512 (2001).

%\bibitem{canieporci}

%\bibitem{doniach} A. M. Black-Schaffer and S. Doniach, \prb {\bf 75}, 134512 (2007).

%

\end{thebibliography}
\end{document}